\documentclass[]{llncs}
\usepackage{lmodern}
\usepackage{amssymb,amsmath}
\usepackage{ifxetex,ifluatex}
\usepackage{fixltx2e} % provides \textsubscript
\ifnum 0\ifxetex 1\fi\ifluatex 1\fi=0 % if pdftex
  \usepackage[T1]{fontenc}
  \usepackage[utf8]{inputenc}
\else % if luatex or xelatex
  \ifxetex
    \usepackage{mathspec}
    \usepackage{xltxtra,xunicode}
  \else
    \usepackage{fontspec}
  \fi
  \defaultfontfeatures{Mapping=tex-text,Scale=MatchLowercase}
  
\fi
\IfFileExists{upquote.sty}{\usepackage{upquote}}{}
\IfFileExists{microtype.sty}{%
\usepackage{microtype}
\UseMicrotypeSet[protrusion]{basicmath} % disable protrusion for tt fonts
}{}
\ifxetex
  \usepackage[setpagesize=false, % page size defined by xetex
              unicode=false, % unicode breaks when used with xetex
              xetex]{hyperref}
\else
  \usepackage[unicode=true]{hyperref}
\fi
\hypersetup{breaklinks=true,
            bookmarks=true,
            pdfauthor={Michael Färber ; Chad Brown},
            pdftitle={Internal Guidance for Satallax},
            colorlinks=false,
            citecolor=blue,
            urlcolor=blue,
            linkcolor=magenta,
            pdfborder={0 0 0}}
\usepackage[numbers]{natbib}
\usepackage{longtable,booktabs}
\providecommand{\tightlist}{%
  \setlength{\itemsep}{0pt}\setlength{\parskip}{0pt}}
\title{Internal Guidance for Satallax}
\author{Michael Färber \inst{1} \and Chad Brown \inst{2}}
\titlerunning{Internal Guidance for Satallax}
\authorrunning{M. Färber and C. Brown}
\institute{Universität Innsbruck, Austria
\linebreak \email{\href{mailto:michael.faerber@uibk.ac.at}{michael.faerber@uibk.ac.at}} \and Czech Technical University in Prague, Czech Republic}
\date{}
\usepackage{centerfloat}
\usepackage{operators}
\usepackage{pgfplots}
\usepgfplotslibrary{units}
\usepackage[svgpath=graphics/]{svg}

\ifx\paragraph\undefined\else
\let\oldparagraph\paragraph
\renewcommand{\paragraph}[1]{\oldparagraph{#1}\mbox{}}
\fi
\ifx\subparagraph\undefined\else
\let\oldsubparagraph\subparagraph
\renewcommand{\subparagraph}[1]{\oldsubparagraph{#1}\mbox{}}
\fi

\begin{document}
\maketitle
\begin{abstract}
We propose a new internal guidance method for automated theorem provers
based on the given-clause algorithm. Our method influences the choice of
unprocessed clauses using positive and negative examples from previous
proofs. To this end, we present an efficient scheme for Naive Bayesian
classification by generalising label occurrences to types with monoid
structure. This makes it possible to extend existing fast classifiers,
which consider only positive examples, with negative ones. We implement
the method in the higher-order logic prover Satallax, where we modify
the delay with which propositions are processed. We evaluated our method
on a simply-typed higher-order logic version of the Flyspeck project,
where it solves 26\% more problems than Satallax without internal
guidance.
\end{abstract}

\section{Introduction}\label{introduction}

Experience can be described as knowing which methods to apply in which
context. It is a result of experiments, which can show a method to
either fail or succeed in a certain situation. Mathematicians solve
problems by experience. When solving a problem, mathematicians gain
experience, which in the future can help them to solve harder problems
that they would not have been able to solve without the experience
gained before.

Fully automated theorem provers (ATPs) attempt to prove mathematical
problems without user interaction. A thriving field of research is how
to make ATPs behave more like mathematicians, by learning which
decisions to take from previous proof attempts, in order to find more
proofs in shorter time, and to prove problems that were previously out
of reach for the ATP. Machine learning can help advance that field, for
it provides techniques to model experience and to compare the quality of
possible decisions. Machine learning approaches to improve ATP
performance include:

\begin{itemize}
\tightlist
\item
  \textbf{Premise selection}: Preselecting a set of axioms for a problem
  can be done as a preprocessing step or inside the ATP at the beginning
  of proof search. Examples of this technique are the Sumo INference
  Engine (SInE) \citep{hoder2011-sine} and E.T. \citep{kaliszyk2015-et}.
\item
  \textbf{Internal guidance}: Unlike premise selection, internal
  guidance influences choices made during the proof search. The
  \emph{hints} technique \citep{veroff1996-hints} was among the earliest
  attempts to directly influence proof search by learning from previous
  proofs. Other systems are E/TSM \citep{schulz2000-phd}, an extension
  of E \citep{schulz2013-e} with term space maps, and MaLeCoP
  \citep{urban2011-malecop} respectively FEMaLeCoP
  \citep{kaliszyk2015-femalecop}, which are versions of leanCoP
  \citep{otten2008-leancop} extended by Naive Bayesian learning.
\item
  \textbf{Learning of strategies}: Finding good settings for ATPs
  automatically has been researched for example in the Blind
  Strategymaker (BliStr) project \citep{urban2015-blistr}.
\item
  \textbf{Learning of strategy choice}: Once one has found good ATP
  strategies for different sets of problems, it is not directly clear
  which strategies to apply for which time when encountering a new
  problem. This problem was treated in the Machine Learning of
  Strategies (MaLeS) \citep{kuehlwein2014-phd}.
\end{itemize}

In this paper, we show an internal guidance algorithm for ATPs that use
(variations of) the given-clause algorithm. Specifically, we study a
Naive Bayesian classification method, introduced for the connection
calculus in FEMaLeCoP, and generalise it by measuring label occurrences
with an arbitrary type having monoid structure, in place of a single
number. This generalisation has the benefit that it can handle positive
and negative occurrences. As a proof of concept, we implement the
algorithm in the ATP Satallax \citep{brown2012-satallax}, using no
features at all, which already solves 26\% more problems given the same
amount of time, and which can solve about as many problems in 1s than
without internal guidance in 2s.

\section{Naive Bayesian classifier with monoids}\label{sec:naivebayes}

\subsection{Motivation}\label{motivation}

Many automated theorem provers have a proof state in which they make
decisions, by ranking available choices (e.g.~which proposition to
process) and choosing the best one. This is related to the
classification problem in machine learning, which takes data about
previous decisions, i.e.~which situation has led to which choice, and
then orders choices by usefulness for the current situation.

For example, let us assume that the state of the theorem prover is
modelled by the set of constants appearing in the previously processed
propositions or in the conjecture. Let our conjecture be
\(x + y = y + x\) and let our premises include

\begin{align}
  \forall P. [ P(0) \implies (\forall x. P(x) \implies P(s(x)))
    \implies \forall x. P(x) ], \label{eq:num_induct} \\
  x + 0 = x. \label{eq:plus_0}
\end{align}

If we first process \autoref{eq:num_induct}, the prover state is
characterised by \(F = \{ +, s, 0 \}\). If we then continue to process
\autoref{eq:plus_0} and it turns out that this contributes to the final
proof, we register that in the situation \(F\), \autoref{eq:plus_0} was
useful.

In other proof searches, processing \autoref{eq:plus_0} in a certain
prover state will not contribute towards the final proof. We call such
situations negative examples.

Intuitively, we would like to apply propositions in situations that are
similar to those in which the propositions were useful, and avoid
processing propositions in situations similar to those where the
propositions were useless. In general, examples (positive and negative)
can be characterised by a prover state \(F\) and a proposition \(l\)
that was processed in state \(F\). This makes it possible to treat the
choice of propositions as classification problem. In the next section,
we show how to rank choices based on previous experience.

\subsection{Classifiers with positive
examples}\label{classifiers-with-positive-examples}

A classifier takes pairs \((F, l)\), relating a set of features \(F\)
with a label \(l\), and produces a function that, given a set of
features, predicts a label. Classifiers can be characterised by a
function \(r(l, F)\), which represents the relevance of a label wrt a
set of features. For internal guidance, we use \(r\) to estimate the
relevance of a clause \(l\) to process in the current prover state
\(F\).

A Bayesian classifier estimates the relevance of a label by its
probability to occur with a set of features, i.e. \(P(l \mid F)\). By
using the Naive Bayesian assumption that features are conditionally
independent, the conditional probability is: \[ P(l \mid F)
 = \frac{P(l) P(F \mid l)}{P(F)}
 = \frac{P(l) \prod_{f \in F} P(f \mid l)}{P(F)}
 \varpropto P(l) \prod_{f \in F} P(f \mid l).
\] To increase numerical stability, we use sums of logarithms.
Furthermore, we weight the probabilities with the inverse document
frequency (IDF) of the features, and we omit the constant factor
\(P(F)\). The resulting classifier then is:
\[ r(l, F) = \log P(l) + \sum_{f \in F} \log (\idf(f_i)) \log P(f \mid l). \]
In FEMaLeCoP, the simplified probability functions\footnote{We omitted
  several constant factors. Furthermore, FEMaLeCoP considers also
  features of training examples that are \emph{not} part of the features
  \(F\), albeit this is a further derivation of the theoretical model.}
are approximated by
\[P(l) \approx D_l, \qquad P(f \mid l) \approx \begin{cases}
  c & \text{if } D_{l,f} = 0 \\
  \frac{D_{l,f}}{D_l} & \text{otherwise}
  \end{cases}\] where \(D_{l,f}\) denotes the number of times \(l\)
appeared among the training examples in conjunction with \(f\), \(D_l\)
denotes how often \(l\) appeared among all training examples, and \(c\)
is a constant.

\subsection{Generalised classifiers}\label{generalised-classifiers}

In our experiments, we found negative training examples to be crucial
for internal guidance. Therefore, we generalised the classifier to
represent the type of occurrences as a \emph{commutative monoid}.

\begin{definition}
A pair (M, +) is a \emph{monoid} if there exists a neutral element $0 \in M$
such that for all $x, y, z \in M$, $(x + y) + z = x + (y + z)$ and
$x + 0 = 0 + x = x$. If furthermore $x + y = y + x$, then the monoid
is \emph{commutative}.
\end{definition}

The generalised classifier is instantiated with a commutative monoid
\((M, +)\) and reads triples \((F, l, o)\), which in addition to
features and label now store the label occurrence \(o \in M\). For
example, if the classifier is to support positive and negative examples,
then one can use the monoid \((\mathbb{N} \times \mathbb{N}, +_2)\),
where the first and second elements of the pair represent the number of
positive respectively negative occurrences, the \(+_2\) operation is
pairwise addition, and the neutral element is \((0,0)\). A triple learnt
by this classifier could be \((F, l, (1, 2))\), meaning that \(l\)
occurs with \(F\) once in a positive and twice in a negative way.
Commutativity imposes that the order in which the classifier is trained
does not matter.

We now formally define \(D_l\) (occurrences of label), \(D_{l,f}\)
(co-occurrences of label with feature) and \(\idf\) (inverse document
frequency):

\begin{align*}
  D_l &= \sum \{ o \mid (F, l', o) \in D, l = l' \}, \\
  D_{l,f} &= \sum \{ o \mid (F, l', o) \in D, l = l', f \in F \}, \\
  \idf(f) &= \frac{|D|}{|\{(F, l', o) \mid (F, l', o) \in D, f \in F\}|}
\end{align*}

With this, our classifier for positive and negative examples can be
defined as follows: \[P(l) = \frac{|p-n|}{p+n}(c_pp + c_nn), \qquad
P(f_i \mid l) = \begin{cases}
  c & \text{if } D_{l,f} = 0 \\
  c_p \frac{p_f}{p} + c_n \frac{n_f}{n} & \text{otherwise}
\end{cases}\] where \((p,n) = D_l\), \((p_f, n_f) = D_{l,f}\), and
\(c\), \(c_p\), and \(c_n\) are constants. The term
\(\frac{|p-n|}{p+n}\) represents \emph{confidence} and models our
intuition that labels which appear always in the same role (say, as
positive example) should have a greater influence than more ambivalent
labels. For example, if a label occurs about the same number of times as
positive and as negative example, confidence is approximately 0, and
when a label is almost exclusively positive or negative, confidence is
1.

We call \(D_l\), \(D_{l,f}\), and \(\idf\) classification data. They are
precalculated to allow fast classification. Furthermore, new training
examples can be added to existing classification data efficiently,
similarly to \citep{kaliszyk2015-femalecop}.

\section{Learning scenarios}\label{sec:scenarios}

In this section, we still consider ATPs as black boxes, taking as input
a problem and classification data for internal guidance, returning as
output training data (empty if the ATP did not find a proof).

\begin{figure}
\subfloat[Online learning.]{\includesvg[width=200pt]{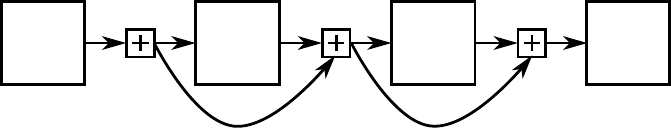}}

\subfloat[Offline learning.]{\includesvg[width=200pt]{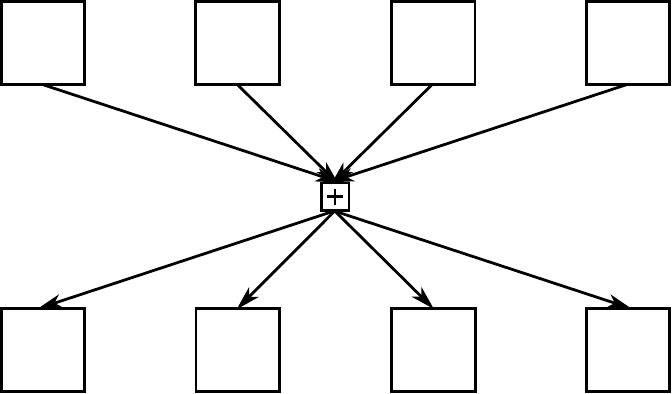}}
\caption{
  Comparison of online and offline learning.
  The large boxes symbolise an ATP proof search, which takes
  classifier data and returns training data (empty if no proof found).
  The small "$+$" boxes combine classifiers and training data, returning
  new classifier data.
  \label{fig:onoffline}
}
\end{figure}

We propose two different scenarios to generate training data and to use
it in subsequent proof searches, see \autoref{fig:onoffline}:

\begin{itemize}
\tightlist
\item
  On-line learning: We run the ATP on every problem with classification
  data. For every problem the ATP solves, we update the classifier with
  the training data from the ATP proof.
\item
  Off-line learning: We first run the ATP on all problems without
  classification data, saving training data for every problem solved. We
  then create classification data from the training data and rerun the
  ATP with the classifier on all problems.
\end{itemize}

While the second scenario can be parallelised, thus taking less
wall-clock time, it has to treat every problem twice in the worst case
(namely when every problem fails), thus taking up to double the CPU time
of the first scenario.

\section{Internal guidance for given-clause
provers}\label{internal-guidance-for-given-clause-provers}

Variants of the given-clause algorithm are commonly used in
refutation-based ATPs, such as Vampire \citep{kovacs2013-vampire} or E
\citep{schulz2013-e}.\footnote{Technically, our reference prover
  Satallax does not implement a given-clause algorithm, as Satallax
  treats terms instead of clauses, and it interleaves the choice of
  unprocessed terms with other commands. However, for the sake of
  internal guidance, we can consider Satallax to implement a version of
  the given-clause algorithm. We describe the differences in more detail
  in \autoref{sec:implementation}.} We introduce a simple version of the
algorithm: Given an initial set of clauses to refute, the set of
\emph{unprocessed} clauses is initialised with the initial set of
clauses, and the set of \emph{processed} clauses is the empty set. At
every iteration of the algorithm, a \emph{given clause} is selected from
the unprocessed clauses and moved to the processed clauses, possibly
generating new clauses which are moved to the unprocessed clauses. The
algorithm terminates as soon as either the set of unprocessed clauses is
empty or the empty clause was generated.

The integration of our internal guidance method into an ATP with
given-clause algorithm involves two tasks: The recording of training
data, and the ranking of unprocessed clauses, which influences the
choice of the given clause. To reduce the amount of data an ATP has to
load for internal guidance, we process training data and transform it
into classification data outside of the ATP. We describe these tasks
below in the order they are executed when no internal guidance data is
present yet.

\subsection{Recording training data}\label{sec:training-data}

Recording training data can be done in different fashions:

\begin{itemize}
\tightlist
\item
  \textbf{In situ}: Information about clause usage is recorded every
  time an unprocessed clause gets processed. This method allows for more
  expressive prover state characterisation, on the other hand, we found
  it to decrease the proof success rate, as the recording of proof data
  makes the inference slower.
\item
  \textbf{Post mortem}: Only when a proof was found, information about
  clause usage is reconstructed. As this method does not place any
  overhead on the proof search, we resorted to post-mortem recording,
  which is still sufficiently expressive for our purposes.
\end{itemize}

For every proof, we save: conjecture (if one was given), axioms \(A\)
(premises given in the problem), processed clauses \(C\), and clauses
\(C_+\) that were used in the final proof (\(C_+ \subseteq C\)). We call
such information for a single proof a \emph{training datum}. We ignore
unprocessed clauses, as we cannot easily estimate whether they might
have contributed to a proof.

\subsection{Postprocessing training data}\label{sec:postprocessing}

In our experiments, we frequently encounter clauses that are the same,
differing only by containing different Skolem constants. To this end, we
process the training data before creating classification data from it.
We tried different techniques to handle Skolem constants, as well as
other postprocessing methods:

\begin{itemize}
\tightlist
\item
  \textbf{Skolem filtering}: We discard clauses containing any Skolem
  constants.
\item
  \textbf{Consistent Skolemisation}: We normalise Skolem constants
  inside all clauses, similarly to \citep{urban2011-malecop}. That is, a
  clause \(P(x,y,x)\), where \(x\) and \(y\) are Skolem constants,
  becomes \(P(c_1,c_2,c_1)\).
\item
  \textbf{Consistent normalisation}: Similarly to consistent
  Skolemisation, we normalise \emph{all} symbols of a clause. That is,
  \(P(x, y, x)\) as above becomes \(c_1(c_2, c_3, c_2)\). This allows
  the ATP to discover similar groups of clauses, for example
  \(a + b = b + a\) and \(a * b = b * a\) both map to
  \(c_1(c_2, c_3) = c_1(c_3, c_2)\), but on the other hand, this also
  maps possibly different clauses such as \(P(x)\) and \(Q(z)\) to the
  same clause. Still, in problem collections which do not share a common
  set of function constants (such as TPTP), this method is suitable.
\item
  \textbf{Inference filtering}: An interesting experiment is to discard
  all clauses generated during proof search that are not part of the
  initial clauses.
\end{itemize}

We denote the consistent Skolemisation/normalisation of a clause \(c\)
described above as \(\mathcal{N}(c)\).

\subsection{Transforming training data to classification
data}\label{transforming-training-data-to-classification-data}

For a given training datum with processed clauses \(C\) and proof
clauses \(C_+\), we define the corresponding classifier data to be:
\[\{(\mathcal{F}(c), c, (1,0)) \mid c \in C_+ \} \cup
  \{(\mathcal{F}(c), c, (0,1)) \mid c \in C \setminus C_+ \},\] where
\(\mathcal{F}(c)\) denotes the features of a clause. We use the monoid
\((\mathbb{N} \times \mathbb{N}, +_2, (0,0))\) introduced in
\autoref{sec:naivebayes}, storing positive and negative examples. The
classifier data of the whole training data is then the (multiset) union
of the classifier data of the individual training data.

\subsection{Clause ranking}\label{sec:ranking}

This section describes how our internal guidance method influences the
choice of unprocessed clauses using a previously constructed classifier.

At the beginning of proof search, the ATP loads the classifier. Some
learning ATPs, such as E/TSM \citep{schulz2000-phd}, select and prepare
knowledge relevant to the current problem before the proof search.
However, as we store classifier data in a hash table, filtering
irrelevant knowledge to the problem at hand would require a relatively
slow traversal of the whole table, whereas lookup of knowledge is fast
even in the presence of a large number of irrelevant facts. For this
reason we do not filter the classification data per problem.

Then, at every choice point, i.e.~every time the ATP chooses a clause
from the unprocessed clauses \(C\), it picks a clause \(c \in C\) that
maximises the clause rank \(R(c,F)\), where

\[R(c, F) = r_{\text{ATP}}(c) + r(\mathcal{N}(c), F)\]

and:

\begin{itemize}
\tightlist
\item
  \(r_{\text{ATP}}(c)\) is an ATP function that calculates the relevance
  of a clause with traditional means (such as weight, age, \ldots{}),
\item
  \(F\) is the current prover state,
\item
  \(r(c, F)\) is the Naive Bayesian ranking function as shown in
  \autoref{sec:naivebayes}, and
\item
  \(\mathcal{N}(c)\) is the normalisation function as introduced in
  \autoref{sec:postprocessing}.
\end{itemize}

\section{Tuning of guidance
parameters}\label{tuning-of-guidance-parameters}

We employed two different methods to automatically find good parameters
for internal guidance, such as \(c\), \(c_p\), and \(c_n\) from
\autoref{sec:naivebayes}.

\subsection{Off-line tuning}\label{off-line-tuning}

Off-line tuning analyses existing training data and attempts to find
parameters that give proof-relevant clauses from the training data a
high rank, while giving proof-irrelevant clauses a low rank. To do this,
we evaluate for every training datum the following formula, which adds
for every proof-relevant clause the number of proof-irrelevant clauses
that were ranked higher:

\[ \sum_{c_+ \in C_+} |\{ c \mid R(c, F) > R(c_+, F_+), c \in C \setminus C_+ \}|, \]

where \(C\) and \(C_+\) come from the training datum (see
\autoref{sec:training-data}), \(F\) and \(F_+\) are the features of the
prover states when \(c\) respectively \(c_+\) were processed (we
reconstruct these from the training datum), and \(R\) is the ranking
formula from \autoref{sec:ranking}.

In the end, we sum up the results of the formula above for all training
data, and take the guidance parameters which minimise that sum.

\subsection{Particle Swarm Optimisation}\label{sec:pso}

Particle Swarm Optimisation \citep{kennedy1995-pso} (PSO) is a standard
optimisation algorithm that can be applied to minimise the output of a
function \(f(\vec x)\), where \(\vec x\) is a vector of continuous
values. A \emph{particle} is defined by a location \(\vec x\) (a
candidate solution for the optimisation problem) and a velocity
\(\vec v\). Initially, \(p\) particles are created with random locations
and velocities. Then, at every iteration of the algorithm, a new
velocity is calculated for every particle and the particle is moved by
that amount. The new velocity of a particle is: \[ \vec v(t+1) =
    \omega \cdot \vec v(t)
  + \phi_p \cdot \vec r_p \cdot (\vec b_p(t) - \vec x(t))
  + \phi_g \cdot \vec r_g \cdot (\vec b_g(t) - \vec x(t)),
\] where:

\begin{itemize}
\tightlist
\item
  \(\vec v(t)\) is the old velocity of the particle,
\item
  \(\vec b_p(t)\) is the location of the best previously found solution
  among all particles,
\item
  \(\vec b_g(t)\) is the location of the best previously found solution
  of the particle,
\item
  \(\vec r_p\) and \(\vec r_g\) are random vectors generated at every
  evaluation of the formula, and
\item
  \(\omega = 0.4\), \(\phi_p = 0.4\), and \(\phi_g = 3.6\) are
  constants.
\end{itemize}

We apply PSO to optimise the performance of an ATP on a problem set
\(S\). For this, we define \(f(\vec x)\) to be the number of problems in
\(S\) the ATP can solve with a set of flags being set to \(\vec x\) and
with timeout \(t\). We then run PSO and take the best global solution
obtained after \(n\) iterations. We fixed \(t = 1s\), \(p = 300\), and
\(|S| = 1000\). The algorithm has worst-case execution time
\(t \cdot p \cdot n \cdot |S|\).

\section{Implementation}\label{sec:implementation}

We implement our internal guidance in Satallax version 2.8. Satallax is
an automated theorem prover for higher-order logic, based on a tableaux
calculus with extensionality and choice. It is written in OCaml by Brown
\citep{brown2012-satallax}. Satallax implements a priority queue, on
which it places several kinds of proof search commands: Among the 11
different commands in Satallax 2.8, there are for example proposition
processing, mating, and confrontation. Proof search works by processing
the commands on the priority queue by descending priority, until a proof
is found or a timeout is reached. The priorities assigned to these
commands are determined by \emph{flags}, which are the settings Satallax
uses for proof search. A set of flag settings is called a \emph{mode}
(in other ATPs frequently called \emph{strategies}) and can be chosen by
the user upon the start of Satallax. Similar to other modern ATPs such
as Vampire \citep{kovacs2013-vampire} or E \citep{schulz2013-e},
Satallax also supports timeslicing via \emph{strategies} (in other ATPs
frequently called \emph{schedules}), which define a set of modes
together with time amounts Satallax calls each mode with. Formally, a
strategy is a sequence \([(m_1, t_1), \dots, (m_n, t_n)]\), where
\(m_i\) is a mode and \(t_i\) the time to run the mode with. The total
time of the strategy is the sum of times, i.e.
\(t_{\Sigma}(S) = \sum_{(m, t) \in S} t\).

As a side-effect of this work, we have extended Satallax with the
capability of loading user-defined strategies, which was previously not
possible as strategies were hard-coded into the program. Furthermore, we
implemented modifying flags via the command line, which is useful
e.g.~to change a flag among all modes of a strategy without changing the
flag among all files of a strategy. We used this extensively in the
automatic evaluation of flag settings via PSO, as shown in
\autoref{sec:pso}.

When running Satallax with a strategy \(S\) and a timeout \(t_{max}\),
then all the times of the strategy are multiplied by
\(\frac{t_{max}}{t_{\Sigma}(S)}\) if \(t_{max} > t_{\Sigma}(S)\), to
divide the time between modes appropriately when running Satallax for
longer than what the strategy \(S\) specifies. Then, every mode \(m_i\)
in the strategy is run sequentially for time \(t_i\) until a proof is
found or the timeout \(t_{max}\) is hit.

An analysis of several proof searches yielded that on average, more than
90\% of commands put onto the priority queue of Satallax are proposition
processing commands, which correspond to processing a clause from the
set of unprocessed clauses in given-clause provers. For that reason, we
decided to influence the priority of proposition processing commands,
giving those propositions with a high probability of being useful a
higher priority. The procedure follows the one described in
\autoref{sec:ranking}, but the ranking of a proposition is performed
when the proposition processing command is put onto the priority queue,
and the Naive Bayes rank is added to the priority that Satallax without
internal guidance would have assigned to the command. As other types of
commands are in the priority queue as well, we pay attention not to
influence the priority of term processing commands too much (by choosing
too large guidance parameters), as this can lead to disproportionate
displacement of other commands.

To record training data, we use the terms from the proof search that
contributed to the final proof. For this, Satallax uses \texttt{picomus}
\citep{biere2008-picosat} to construct a minimal unsatisfiable core.

To characterise the prover state of Satallax, we tried different kinds
of features:

\begin{itemize}
\tightlist
\item
  Symbols of processed terms: We collect the symbols of all processed
  propositions at the time a proposition is inserted into the priority
  queue and call these symbols the features of the proposition. However,
  this experimentally turned out to be a bad choice, because the set of
  features for each proposition grows quite rapidly, as the set of
  processed propositions grows monotonically.
\item
  Axioms of the problem: We associate every proposition processed in a
  proof search with all the axioms of the problem. In contrast to the
  method above, this associates the same features to all propositions
  processed during the proof search for a problem, and is thus more a
  characterisation of the problem (similar to TPTP characteristics
  \citep{sutcliffe2010-thf}) than of the prover state.
\end{itemize}

In our experiments, just calculating the influence of these features
without them actually influencing the priority makes Satallax prove less
problems (due to the additional calculation time), and the positive
impact of the features on the proof search does not compensate for the
initial loss of problems. Therefore, we currently do not use features at
all and associate the empty set of features to all labels, i.e.
\(\mathcal{F}(c) = \{\}\). However, it turns out that even without
features, learning from previous proofs can be quite effective, as shown
in the next section.

\section{Evaluation}\label{evaluation}

To evaluate the performance of our internal guidance method in Satallax,
we used a THF0 \citep{sutcliffe2010-thf} version (simply-typed
higher-order logic) of the top-level theorems of the Flyspeck
\citep{hales2015-flyspeck} project, as generated by Kaliszyk and Urban
\citep{kaliszyk2014-flyspeck}. The test set consists of 14185 problems
from topology, geometry, integration, and other fields. The premises of
each problem are the actual premises that were used in the Flyspeck
proofs, amounting to an average of 84.3 premises per problem.\footnote{The
  test set, as well as our modified version of Satallax and instructions
  to recreate our evaluation, can be found under:
  \url{http://cl-informatik.uibk.ac.at/~mfaerber/satallax.html}.} We
used an Intel Core i3-5010U CPU (2.1 GHz Dual Core, 3 MB Cache) and ran
maximally one instance of Satallax at a time.

To evaluate the performance of the off-line learning scenario described
in \autoref{sec:scenarios}, we run Satallax on all Flyspeck problems,
generating training data whenever Satallax finds a proof. We use the
Satallax 2.5 strategy (abbreviated as ``S2.5''), because the newest
strategy in Satallax 2.8 can not always retrieve the terms that were
used in the final proof, which is important to obtain training data.

As the off-line learning scenario involves evaluating every problem
twice (once to generate training data and once to prove the problem with
internal guidance), it doubles runtime in the worst case, i.e.~if no
problem is solved. Therefore, a user might like to compare its
performance to simply running the ATP with double timeout directly: When
increasing the timeout from 1s to 2s, the number of solved problems
increases from 2717 to 3394. However, this is mostly due to the fact
that Satallax tries more modes, so to measure the gain in solved
problems more fairly, we create a strategy ``S2.5\_1s'' which contains
only those modes that were already used during the 1s run, and let each
of them run about double the time. This strategy proves 2845 problems in
2s.

We now compare the different postprocessing options introduced in
\autoref{sec:postprocessing}. For this, we create a classifier from the
training data gathered during the 1s run. We then run Satallax with
internal guidance in off-line learning mode with 1s timeout and the
Satallax 2.5 strategy. We perform this procedure for each postprocessing
option. We call a problem ``lost'' that Satallax with guidance could not
solve and Satallax without guidance could. Vice versa for ``gained''.
The results are given in \autoref{tab:guidance-offline}. We perform best
when influencing only the priority of axioms (inference filtering),
solving 786 problems that could not be solved by Satallax in 1s without
internal guidance.

\begin{longtable}[c]{@{}lrrr@{}}
\caption{Comparison of postprocessing options.
\label{tab:guidance-offline}}\tabularnewline
\toprule
Postprocessing & Solved & Lost & Gained\tabularnewline
\midrule
\endfirsthead
\toprule
Postprocessing & Solved & Lost & Gained\tabularnewline
\midrule
\endhead
Consistent normalisation & 1911 & 920 & 114\tabularnewline
Consistent Skolemisation & 1939 & 885 & 107\tabularnewline
None & 2166 & 688 & 137\tabularnewline
Skolem filtering & 3395 & 98 & 776\tabularnewline
Inference filtering & 3428 & 75 & 786\tabularnewline
\bottomrule
\end{longtable}

To evaluate online learning, we run Satallax on all Flyspeck problems by
ascending order, accumulating training data and using it for all
subsequent proof searches. We filter away terms in the training data
that contain Skolem variables. As result, Satallax with online learning,
running 1s per problem, solves 3374 problems (59 lost, 716 gained),
which is a plus of 24\%.

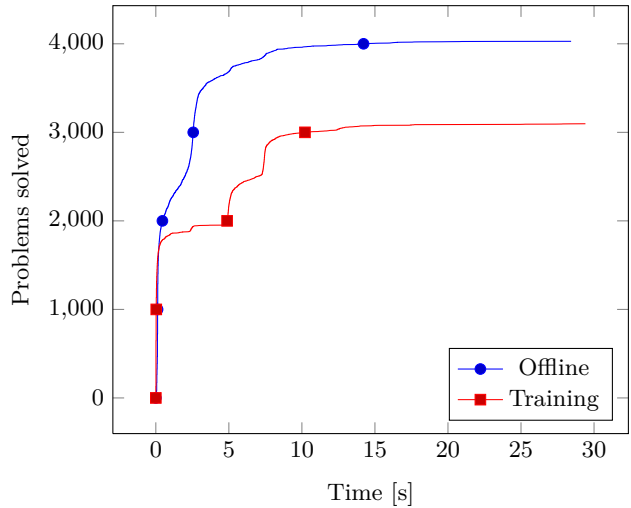
\begin{figure}
\begin{tikzpicture}
\begin{axis}
[ legend pos=south east
, xlabel=Time
, x unit=s
, ylabel=Problems solved
, mark repeat={1000}
]
\addplot table {data/offline.integrated};
\addlegendentry{Offline};
\addplot table {data/training.integrated};
\addlegendentry{Training};
\end{axis}
\end{tikzpicture}
\caption{Problems solved in a certain time.}
\label{fig:30s}
\end{figure}

In the next experiment, we evaluate the prover performance with the
``S2.5\_1s'' strategy and a timeout of 30s. For this, we use an 48-core
server with 2.2GHz AMD Opteron CPUs and 320GB RAM, running 10 instances
of Satallax in parallel. First, we run Satallax without internal
guidance for 30s, which solves 3097 problems. Next, we create from the
training data a classifier with Skolem filtering, which takes 3s and
results in a 1.8M file. Finally, we run Satallax with internal guidance
in off-line learning mode using the classifier. This proves 4028
problems in 30s, which is a plus of 30\%. Results are shown in
\autoref{fig:30s}. The ``jumps'' in the data stem from changes of modes.

\section{Conclusion}\label{conclusion}

We have shown how to integrate internal guidance into ATPs based on the
given-clause algorithm, using positive as well as negative examples. We
have demonstrated the usefulness of this method experimentally, showing
that on a given test set, we can solve up to 26\% more problems. ATPs
with internal guidance could be integrated into hammer systems such as
Sledgehammer (which can already reconstruct Satallax proofs
\citep{sultana2013-sledgehammer}) or HOL(y)Hammer
\citep{kaliszyk2014-flyspeck}, continually improving their success rate
with minimal overhead. It could also be interesting to learn internal
guidance for ATPs from subgoals given by the user in previous proofs.
Currently, we learn only from problems we could find a proof for, but in
the future, we could benefit from considering also proof searches that
did not yield proofs. Furthermore, it would be interesting to see the
effect of negative examples on existing ATPs with internal guidance,
such as FEMaLeCoP. We believe that finding good features that
characterise prover state are important to further improve the learning
results.

\section*{Acknowledgements}\label{acknowledgements}
\addcontentsline{toc}{section}{Acknowledgements}

We would like to thank Sebastian Joosten and Cezary Kaliszyk for reading
initial drafts of the paper, and especially Josef Urban for inspiring
discussions and inviting the authors to Prague. Furthermore, we would
like to thank the anonymous IJCAR referees for their valuable comments.

This work has been supported by the Austrian Science Fund (FWF) grant
P26201 as well as by the European Research Council (ERC) grant
AI4REASON.

\renewcommand\refname{References}
\bibliography{literature}

\end{document}